# The Polarized X-ray Universe: Insights and Discoveries

Arbind Pradhan[1], Sree Bhattacherjee[1], Biplob Sarkar[1]
[1]Department of Applied Sciences, Tezpur University, Tezpur, Assam-784028
*Email ID: app22111@tezu.ac.in (Corresponding Author)

**Abstract**: Polarization is one of the fundamental natures of electromagnetic radiation. The detection of polarization or polarized photons from distant X-ray radiating systems (such as X-ray binaries (XBs), active galactic nuclei (AGN), pulsars, and stars) complements the timing, spectral, and imagining analysis to better understand the physical mechanisms taking place in these sources. Polarization has enhanced the understanding of the internal geometry of these systems and their vicinity. Polarized X-rays can be generated either directly through non-thermal physical processes in the presence of a magnetic field(**B**) or through the scattering of unpolarized thermal radiation within plasma structures such as an accretion disk. X-ray polarization can measure the two important independent parameters, the polarization degree (PD) and polarization angle (PA) of the X-ray photons. These parameters are crucial as they reveal the characteristics of particles in such a strong magnetic and gravitational field. In this chapter, we have discussed (i) the basic idea of polarization, (ii) some distant sources radiating polarized X-ray photons, (iii) missions dedicated to observing polarized X-ray photons, and (iv) recent breakthroughs and upcoming missions.

**Keywords**: Black hole, Neutron star, Polarimetry, X-rays, Magnetic fields.

## 1. Introduction

The nature of electromagnetic (EM) radiation emanating from celestial objects conveys insightful knowledge regarding their inherent attributes and ambient conditions. One of the basic characteristics of EM radiation is its polarization. The polarized light from astronomical objects has helped us better understand these distant objects. The first polarized light observed and studied from an astronomical object is dated from the mid-19th century. Secchi, (1860) considered the reflection by the Moon to cause the linear polarization of sunlight, whereas Edlund, (1860) treated the linear polarization as from the solar corona. With the advancement in the observation techniques of polarized light in different regimes, like optical to radio polarimetry (Wilson et al., 2013) in the 1940s and in the 1970s, X-ray polarimetry (Weisskopf et al., 1978) has facilitated the study of astronomical distant objects.

X-ray astronomy has enhanced the understanding of accretion processes, radiative processes, the geometry of the compact object (CO) or system consisting of a CO and its close vicinity, and the system's extreme gravitational and magnetic field (**B**). The investigation of X-rays from celestial objects using techniques such as spectroscopy, timing or temporal studies, and imaging has advanced over the years, giving more insight into these distant sources. X-ray polarization has evolved as the fourth pillar to get more insight into this system. The launch of X-ray polarization dedicated missions has provided much new information and opened many questions. There are many important standing questions that are expected to be addressed by X-ray polarization studies, such as determining the geometry and physical properties of accretion flows and coronas in XBs. These studies can also provide a framework for testing general relativity, understanding relativistic jets in systems like blazars and microquasars, and elucidating the magnetic field structure in these systems. Furthermore, polarization studies can distinguish between different emission models and various coronal geometries, such as slab-like, spherical, or patchy.

This chapter gives a basic idea of X-ray polarization, some distant astronomical objects radiating polarized X-ray photons, missions dedicated to X-ray polarization, and recent breakthroughs and upcoming missions.

## 2. Basic idea of polarization.

We know that the EM waves are transverse in nature, where the electric field (**E**) oscillates in one plane, and perpendicular to that, the **B** oscillates. Further, they are perpendicular to the wave propagation plane. From Maxwell's equations in vacuum, we consider **E** of EM waves traveling in the z-direction with a speed of light 'c', hence.

$$E(t,z) = E(0,0) \cos(\omega t - kz - \phi) \quad \text{----(1)}$$

Where $\phi$ signifies an arbitrary phase, $k (= \omega/c)$ represents the wave vector, $\omega$ indicates the angular frequency and t denotes time. Decomposing $E(t,z)$ as $E_x$ (x component) and $E_y$ (y component),

$$E_x(t) = E_x(0) \cos(\omega t - \phi_1) \quad \text{----(2)}$$
$$E_y(t) = E_y(0) \cos(\omega t - \phi_2) \quad \text{----(3)}$$

we are considering $z = 0$ for simplicity, and two arbitrary phases $\phi_1$ and $\phi_2$. The angle between the **E**(t) and the positive x-axis is known as the PA. The components like $E_x(0)$, $E_y(0)$, $\phi_1$, and $\phi_2$ with the relative values will give the polarization of the wave (Rybicki & Lightman, 1985; Goldstein, 2017). Basically, there are three types of polarization,

i. Linear Polarization: $\phi_1 = \phi_2$ and PA is constant.
ii. Elliptical Polarization: $\phi_1$ not equal to $\phi_2$, the tip of **E** traces an ellipse in xy plane. The PA is the angle formed by the semi-major axis and the positive x-axis of the ellipse.
iii. Circular Polarization: It represents a specific case of elliptical polarization where $\phi_2 = \phi_1 \pm \pi/2$ and $E_x(0) = E_y(0)$. In the xy-plane, the **E**'s traces a circular path with an angular frequency '$\omega$'.

These types of polarization refer to microscopic polarization as it deals with the individual waves, whereas in the case of astronomical observation, this microscopic polarization is not sensitive. Microscopic polarization occurs when the **E** from the radiation source or the interstellar medium properties prefers a certain orientation (Trippe, 2014). For such cases, the light so observed is partially polarized with a PD, given as,

$$PD = \frac{I_p}{I} (100\%) \quad \text{----(4)}$$

here, '$I_P$' represents the intensity of polarized light, while 'I' signifies the total intensity of the light. The relationship between 'I' and the amplitude of the electric field is expressed as $I \propto E^2$.

### 2.1 Stokes parameters (SPs).

SPs are the sets of four parameters that describe the state of polarization of EM radiation. They are crucial parameters to determine the complete properties of the radiation. They are denoted as I, Q, U, and V. Here, I represents wave's intensity, and Q signifies the difference between the intensities of horizontally ($E_x$) and vertically ($E_y$) polarized light, providing information about linear polarization. U signifies the difference between intensities polarized by the two field components diagonally at 45° and 135°; it also helps in probing linear polarization. Finally, V represents circularly polarized intensity. In astronomical observations, it is convenient to represent polarization as the intensities rather than the amplitude; hence, the SP provides a better polarization measurement. SPs can be given as,

$$I = \langle E_x^2 \rangle + \langle E_y^2 \rangle$$

$$Q = \langle E_x^2 \rangle - \langle E_y^2 \rangle$$

$$U = 2\langle E_x E_y \cos\delta \rangle$$

$$V = 2\langle E_x E_y \sin\delta \rangle$$

(Stokes, 1851). <..> represents the time average of the enclosed parameter taken over times much larger than $2\pi/\omega$ (Goldstein, 2017; Wilson et al., 2013).

For microscopic polarization, SPs can be related as,

$$I^2 = Q^2 + U^2 + V^2$$

while for macroscopic polarization, in the case of astronomical observation, the SP can be given as,

$$I_p^2 = Q^2 + U^2 + V^2$$

where $I_p$ is polarized intensity and $I_p \leq I$. Considering equation (4), we get PD and PA as,

$$PD = \sqrt{Q^2 + U^2}/I \quad \text{-----------------------------(5)}$$

$$PA = \frac{1}{2} atan_2 \frac{U}{Q} \quad \text{-----------------------------(6)}$$

, 'atan$_2$' represents the quadrant-preserving arc tangent. X-rays are also transverse EM radiation, so all the above polarization concepts are applied in the X-ray polarization study. The two important measurements that are measured with the X-ray polarization are the determination of PD and PA, which characterize the incoming X-ray photons from the X-ray source. These parameters help to investigate the dynamics of matter and radiation in the presence of strong **B** and gravitational fields. For more extensive information related to X-ray polarization's scientific importance, one can refer to Soffitta, (1997); Lei et al., (1997); Krawczynski et al., (2011).

## 3. Some astronomical objects that emit polarized X-rays.

In general, the polarization of light is primarily a result of reflection and scattering (absorption and re-emission). In this section, we will briefly discuss some astronomical objects radiating in X-rays from which X-ray polarization observation is possible and can give significant insight into the source and its surroundings.

■ Solar Flares:

Sunlight is considered to be an unpolarized light. However, the polarization of sunlight is used to verify the **B**'s direction and **B**'s magnitude of the Sun (West & Smith, 1994). The solar flares are one of the extremely powerful events in the sun. It happens as the electrons get accelerated into the chromosphere by magnetic reconnection in the Sun's corona. It is believed that thermal heating at the reconnection site will cause the soft X-rays to become unpolarized. According to the energy range, X-ray photons can be classified into two categories. For X-rays in the energy range of 12 keV to 120 keV, high-energy photons (with short wavelengths) are referred to as "hard X-rays," whereas "soft X-rays" have lower energies (longer wavelengths) ranging from 0.12 keV to 12 keV (Attwood, 1999). However, electron distribution is anisotropic in nature, it might have a low PD (Emslie & Brown, 1980).

The quantity of photon flux in X-rays emitted from solar flares makes it one of the potential candidates to perform X-ray polarimetry, especially in hard X-rays. It is expected that the hard X-rays will originate from high-energy electron's non-thermal Bremsstrahlung emission. As a result, they are thought to be highly polarized, with the PD varying depending on the electron beam, **B** structure, and photon backscattering from the photosphere (Jeffrey & Kontar, 2011).

■ Binary black Hole Systems:

Black hole binaries (BHB) or black hole X-ray binaries (BHXBs) are systems accreting mass from its companion and emitting polarized X-rays. The lower energy flux (mainly emission from the disk) is dominated by thermal emissions, which are believed to originate from the scattering from the disk atmosphere. Hence, the lower energy polarized X-rays from this system are linked with disk scattering, and the PD varies with system inclination. Depending on the optical depth, the PA may appear parallel or perpendicular to the disk axis. General relativistic phenomena like aberration and beaming, gravito-magnetic frame-dragging, and gravitational lensing result in energy-dependent polarization fractions and degrees (Connors & Stark, 1977; Connors et al., 1980). The "returning radiation"; causes the polarization signature to shift from a horizontal orientation at low energy (perpendicular to the rotation axis, as in Chandrasekhar's flat-space Newtonian limit) to a vertical orientation (parallel to the rotation axis) at higher energies (Krawczynski et al., 2011). The "returning radiation" are X-rays that escape the accretion disk, strike it once more after being redirected by gravity around the BH, and then scatter in the observer's direction.

The probing of hard X-ray polarization can give many insights into the BH and its surroundings. The temperature in the proximity of the BH (disk's inner edge) is very high, and the high-energy photons are generally emitted from this region. The polarization so observed is energy dependent, and that is dependent on optical depth; it will travel smoothly from parallel to the disk to perpendicular or vice versa (Chattopadhyay, 2021). Schnittman & Krolik, (2010) have studied the various coronal geometries by investigating polarization.

■ Neutron Stars (NS) and pulsars:

The X-ray polarization can give better insights into the geometry of the NS system, which includes magnetars, pulsars, and accreting pulsars, as well as the emission processes associated with the NS system.

It also gives information about strong **B**'s and the behavior of matters in them (Weiskopf et al., 2006). There is a debate about the rotating rotation-powered pulsars' emission mechanisms and emission sites. This debate is related to various models, including the polar cap model, which posits that pulsar radiation originates from the polar cap (Daugherty & Harding, 1982); the outer gap model, which suggests that the source lies within the outer magnetosphere (Cheng et al., 2000), and the slot gap model, which suggests that emissions occur from the polar cap to the light cylinder along the last open field line (Dyks & Rudak, 2003), demonstrating unique phase-dependent polarization characteristics. Phase-resolved polarimetry can be employed to assess these models and deepen our understanding of the emission mechanisms and locations in isolated X-ray pulsars.

In magnetars, the **B** is in the range of $10^{14-15}$ G, which is exceptionally high compared to other objects and is a potential candidate to emit highly polarized X-rays. It has a bright, hard X-rays tail (20 − 100 keV); hence, hard X-ray polarization is one of the best probing techniques to investigate the physical processes and to understand the magnetars and their extremely strong **B** (Chattopadhyay, 2021).

Besides these, objects like AGN, blazars, and gamma-ray bursts (GRBs) are also potential candidates for observing the polarized X-rays to get more insight into these systems. The polarimetric study of X-rays emitted from AGNs can be used to investigate the torus's structure (Goosmann & Matt, 2011). The X-ray polarization can provide information on the GRB jet's structure and the mechanism underlying its prompt emission (Granot & Königl, 2003; Eichler & Levinson, 2003; Waxman, 2003). Several missions have been launched in recent years to detect and monitor polarized X-rays from these sources; some are discussed in Section 4.

## 4. Missions dedicated to observing polarized X-ray photons:

The foundation work of the X-ray polarization was started in the late 1960s with Robert Novick's team at Columbia University, which pioneered the use of Bragg diffraction at 45° to detect X-ray polarization. This discovery made it possible to determine the Crab Nebula's polarization initially with sounding rockets and then with the 8th Orbiting Solar Observatory (OSO-8) satellite in the late 1970s. Other early missions contributed to the history of X-ray polarimetry, such as Ariel-5, while the sensitivity of the instrument was very high (Costa, 2024). For several decades, the focus shifted toward improving technologies capable of achieving higher sensitivity and resolution. As a result, the first X-ray polarimetry focused mission, the Imaging X-ray Polarimetry Explorer (IXPE) was launched by NASA marking a major milestone in X-ray polarization study, as it combines advanced detectors with X-ray optics, enabling unprecedented sensitivity and angular resolution.

### 4.1. IXPE:

IXPE is the first entirely dedicated X-ray polarimetry mission to leverage the polarization properties of light emitted by astrophysical sources to enhance our knowledge of X-ray generation in phenomena like NSs, pulsars, and BHs. NASA launched IXPE on a SpaceX Falcon 9 rocket from the Kennedy Space Center, Florida on December 9, 2021. IXPE is functional over the 2-8 keV energy range. It has a good angular resolution of ~30 arcsecond half-power diameter (system), and a sensitive area of 15 mm×15 mm. The IXPE payload (O'Dell et al., 2019;Soffitta et al., 2020;Soffitta et al., 2021) is made up of 3 similar telescope systems that are co-aligned with the spacecraft's pointing axis and equipped with star trackers, an essential component to maintain the attitude determination. Each system functions independently and focuses X-rays onto the corresponding detector units (DUs) using a mirror module assembly (MMA), having a focal length of 4 m (Ramsey et al., 2020). A deployable boom is used to obtain focal length; to maintain a thermal environment consistency, a thermal shock is placed over the boom. A single imaging detector with independent electronics, that is polarization-sensitive, is hosted in each DU. The three DUs and its detector service units, are collectively referred to as the IXPE instrument (Soffitta et al., 2021). As previously stated, the DUs are mechanically fixed to the spacecraft's top deck and rotate 120° with respect to the beam axis. During scientific observations, a metrology system monitors motion between the two ends of the Observatory. This system utilizes a camera mounted on the underside of the Metallic Mirror Module Support Structure (MMSS), which includes the fixed X-ray shield, MMAs and +Z star tracker. For proper pointing and the collection of scientific data, proper alignments between the +Z star tracker, MMAs, and DUs are essential. To endure launch loads, three bipods are attached to the MMSS deck. The system accommodates variations in deployed geometry through a tip/tilt/rotate mechanism, enabling on-orbit adjustments between the top deck-mounted DU and the deployed X-ray optics. Refer to Figure 1 of Soffitta, (2024) for a detailed payload overview.

IXPE has made a significant contribution in X-ray astronomy by pioneering the X-ray polarization study, enabling deeper understanding of the COs about the emission component, orientation and critical insights into their physical and astrophysical properties.

**4.2. X-ray Polarimeter Satellite (XpoSat):**

XpoSat (Paul, 2022) is India's first dedicated polarimetry mission launched by Indian Space Research Organization (ISRO) on January 1, 2024, on a PSLV rocket. The prime objective of this mission is to investigate the dynamics of celestial X-ray sources in extreme conditions. The mission's expected lifetime is five years. The payload of the instrument carries two scientific payloads, i) Polarimeter Instrument in X-rays (POLIX), and ii) X-ray Spectroscopy and Timing (XSPECT). POLIX has been jointly developed by the Raman Research Institute and the U R Rao Satellite Centre (URSC), Bangalore. It is designed for astronomical observations in the 8–30 keV energy band. The apparatus consists of one scatterer, one collimator, and four XR proportional counter detectors positioned around the scatterer. Incoming polarized X-rays are subjected to anisotropic Thomson scattering by the scatterer, composed of low atomic mass material. Since the collimator limits the field of vision to a 3° x 3°, only one light source can be observed at a time for most of the observations. This is the first polarimeter designed exclusively for measurements of polarimetry in the medium X-ray energy band.

XPoSat's payload, XSPECT, is intended to provide excellent spectroscopic resolution and fast timing in soft XRs. It was designed by the Space Astronomy Group URSC/ISRO in Bengaluru. XSPECT enables simultaneous temporal monitoring of soft X-ray emissions in the 0.8–15 keV range, long-term tracking of spectral state changes, and analysis of variations in line flux and profiles. It achieves this by leveraging the extended observation duration required by POLIX to measure X-ray polarization. An array of Swept Charge Devices (SCDs) provides an effective area greater than 30 cm² at 6 keV with an energy resolution of better than 200 eV. By reducing the field of view for XSPECT, passive collimators are utilized to decrease background noise. A variety of sources will be observed by XSPECT, such as low **B** NS Low mass XB (LMXBs) and BH LMXBs, and X-ray pulsars, AGNs and magnetars. Readers may refer to the official site https://www.isro.gov.in/XPoSat.html for further details about the mission.

**5. Recent scientific breakthroughs:**

In this section we discuss some of the recent studies conducted using IXPE observations, since its archival data is publicly available. While XpoSat observations are not yet available and are currently in the verification stage, IXPE has been instrumental in providing insights into the EM emission mechanisms of LMXBs. The study of some of these systems is now discussed; 4U 1630-47, which is a BH-LMXB, was studied using IXPE data (Rawat et al., 2023a, Rawat et al., 2023b, Kushwaha et al., 2023). The findings reveal that in the energy range of 2–8 keV, the polarization degree (PD) is approximately 8%. Furthermore, the PD exhibits a notable increase with energy, rising from about 6% at ~2 keV to approximately 11% at ~8 keV. The study suggested that the source was in high soft state (SS) during the observation, and the observed high PD could be associated with reflected or direct emission from the accretion disk (Rawat et al., 2023a), while in epoch 2, 2023 March 10-13, in contrast to the high SS, the steep power-law state shows a polarization difference of $4.4\sigma$ and a decrease in polarization fraction with energy band. The polarization behaviour that the authors observe in both states could be associated with the self-irradiation of the disk around the BH, subject to frame-dragging effects (Rawat et al., 2023b). While Kushwaha et al., (2023) provide insights into the geometry of the accretion disk of the source 4U 1630-47, during its 2022 outburst, the authors observed a significant PD of $8.33 \pm 0.17$ % and PA of $17.78° \pm 0.60°$ in the IXPE energy range. From the spectro-polarimetric study, the authors detected line absorption features at ~6 keV and ~7 keV, indicating the presence of disk winds; the spin of the BH is also estimated to be $0.927 \pm 0.001$ ($1\sigma$) from the data analysis. Cyg X-1 is a persistent BH-LMXB; Steiner et al., 2024 performed the polarization study for this source in the SS initially using IXPE observations. A weaker polarization of ~2% is obtained in SS in comparison to the hard state (HS), while PA being constant ~26° in both states. The study explained the polarization in the HS with the multiple scattering in corona while, in SS, this can be explained by reflected emission from the accretion disk, resulting from the strong gravitational lensing and high spin of the inner AD, as observed for the BH- LMXB 4U 1957+115 (Marra et al., 2024). For the source LMC X-3, Majumder et al., (2024); Garg et al., (2024), explained the emission mechanism and the coronal geometry, based on spectro-polarimetric analysis conducted using simultaneous observations from various missions. Veledina et al., (2024), studied the target Cyg X-3 using IXPE observations and they detected a strong energy-independent linear polarization perpendicularly oriented to the radio ejection, indicating a collimated beam of outflow from the system constraining its half-opening angle to $15°$; hence the source can serve as a laboratory to understand the supercritical accretion environment. *AstroSat's* Cadmium Zinc Telluride Imager (CZTI) is capable of performing high-energy (100–380 keV) X-ray polarization measurements (Bhalerao et al., 2017). For instance, Chattopadhyay et al., (2024) utilized CZTI to confirm the intermediate hard state and variable jet components in Cygnus X-1.

The polarization study also interprets the accretion geometry of NS-LMXBs. GS 1826−238 is the first NS-LMXB observed by IXPE (Capitanio et al., 2023). Following this, several other NS-LMXBs were also observed. Notable examples include GX 9+9 (Chatterjee et al., 2023; Ursini et al., 2024) and Cyg X-2 (Farinelli et al., 2023). These studies suggest that the primary source of the polarization signal in such systems originates from the boundary layer at the NS surface. Additionally, reflection from the accretion disk may also contribute to the observed polarization. Additional observations have provided further insights into the polarization properties of NS-LMXBs. For instance, in 4U 1820−303 (Di Marco et al., 2023), the disk geometry of the target was predicted. In 4U 1624−49 (Saade et al., 2024), the authors suggested that the high PD observed could arise not only from the boundary layer and the accretion disk but also from a geometrically thin, extended slab corona. For GX 5−1 (Fabiani et al., 2024), a Z-track source, the authors observed a PD of 3.7%±0.4% in the horizontal branch, which decreased to 1.8%±0.4% in the normal-flaring branch. This confirmed that the PD depends on the source's position in the color-color diagram (a plot of the photon count ratio in the hard to soft energy band) and the hardness-intensity track (a plot of the photon count ratio in the hard energy range versus total intensity).

Polarization studies of AGNs have been successfully conducted using IXPE data. Recent results on various AGNs and blazars have provided valuable insights into their complex emission mechanisms, coronal geometry, and dusty torus structure (see Laha et al., (2025) for a detailed review). Furthermore, polarization studies of Seyfert galaxies, such as MCG-05-23-16 (Marinucci et al., 2022) and the Seyfert 1 galaxy IC 4329A (Ingram et al., 2023), have discussed outflows and coronal geometry.

Apart from the XBs, IXPE has also provided insights about the pulsars, such as SMC X-1 (Forsblom et al., 2024), Crab (Vivekanand, 2024, Gonzalez-Caniulef et al., 2024), Swift J0243.6+6124 (Majumder, Chatterjee, et al., 2024), PG 1553+113 (Middei et al., 2023), Hercules X-1 (Garg et al., 2023); blazars such as PKS 2155-304 (Kouch, Liodakis, Middei, et al., 2024), S4 0954+65 (Kouch, Liodakis, Fenu, et al., 2024).

Further studies and observations are still going on to deepen our understanding about the insights of the compact stellar objects.

## Summary

Although IXPE was originally designed for a two-year mission, it remains active and continues to provide valuable observations, aiding our understanding of compact systems. Soon, XpoSat will also release data that researchers can use to further explore these systems and gain new insights. It is important to highlight that the Chinese Academy of Sciences and universities in China, in collaboration with European institutions, is planning to launch the upcoming polarimetry satellite mission, the Enhanced X-ray Timing and Polarimetry mission (eXTP), by the year 2027. This mission will significantly advance our understanding of magnetars and pulsars.

## Acknowledgments

Author Sree Bhattacherjee expresses gratitude for the financial assistance received through the INSPIRE fellowship (Grant No.: DST/INSPIRE Fellowship/[IF220164]) from the Department of Science and Technology (DST), Ministry of Science and Technology, India. Biplob Sarkar acknowledges the UGC-BSR Start-Up-Grant for the financial support to carry out this work (No.F.30-558/2021 (BSR) dated-16/12/2021). Arbind Pradhan acknowledges Tezpur University for providing the institutional fellowship.

## Bibliography

Attwood, D. (1999). *Soft X-Rays and Extreme Ultraviolet Radiation: Principles and Applications*. Cambridge University Press. https://doi.org/10.1017/CBO9781139164429

Bhalerao, V., Bhattacharya, D., Vibhute, A., Pawar, P., Rao, A. R., Hingar, M. K., Khanna, R., Kutty, A. P. K., Malkar, J. P., Patil, M. H., Arora, Y. K., Sinha, S., Priya, P., Samuel, E., Sreekumar, S., Vinod, P., Mithun, N. P. S., Vadawale, S. V., Vagshette, N., … Subbarao, K. (2017). The Cadmium Zinc Telluride Imager on AstroSat. *Journal of Astrophysics and Astronomy*, *38*(2), 31. https://doi.org/10.1007/s12036-017-9447-8

Capitanio, F., Fabiani, S., Gnarini, A., Ursini, F., Ferrigno, C., Matt, G., Poutanen, J., Cocchi, M., Mikusincova, R., & Farinelli, R. (2023). Polarization Properties of the Weakly Magnetized Neutron Star X-Ray Binary GS 1826–238 in the High Soft State. *The Astrophysical Journal*, *943*(2), 129.

Chatterjee, R., Agrawal, V. K., Jayasurya, K. M., & Katoch, T. (2023). Spectro-polarimetric view of bright atoll source GX 9+9 using IXPE and AstroSat. *Monthly Notices of the Royal Astronomical Society: Letters, Volume 521, Issue 1, Pp.L74-L78*, *521*(1), L74. https://doi.org/10.1093/mnrasl/slad026

Chattopadhyay, T. (2021). Hard X-ray polarimetry—An overview of the method, science drivers, and recent findings. *Journal of Astrophysics and Astronomy*, *42*(2), 106. https://doi.org/10.1007/s12036-021-09769-5

Chattopadhyay, T., Kumar, A., Rao, A. R., Bhargava, Y., Vadawale, S. V., Ratheesh, A., Dewangan, G., Bhattacharya, D., Mithun, N. P. S., & Bhalerao, V. (2024). High Hard X-Ray Polarization in Cygnus X-1 Confined to the Intermediate Hard State: Evidence for a Variable Jet Component. *The Astrophysical Journal*, *960*, L2. https://doi.org/10.3847/2041-8213/ad118d

Cheng, K. S., Ruderman, M., & Zhang, L. (2000). A three-dimensional outer magnetospheric gap model for gamma-ray pulsars: Geometry, pair production, emission morphologies, and phase-resolved spectra. *Astrophysical Journal Letters*. https://hub.hku.hk/handle/10722/43327

Connors, P. A., Piran, T., & Stark, R. F. (1980). Polarization features of X-ray radiation emitted near black holes. *Astrophysical Journal, Part 1, Vol. 235, Jan. 1, 1980, p. 224-244. Research Supported by the Science Research Council and University of Texas*, *235*, 224–244.

Connors, P. A., & Stark, R. F. (1977). Observable gravitational effects on polarised radiation coming from near a black hole. *Nature*, *269*(5624), 128–129. https://doi.org/10.1038/269128a0

Costa, E. (2024). General History of X-Ray Polarimetry in Astrophysics. In *Handbook of X-ray and Gamma-ray Astrophysics* (pp. 5663–5682). Springer. https://link.springer.com/content/pdf/10.1007/978-981-19-6960-7_140.pdf

Daugherty, J. K., & Harding, A. K. (1982). Electromagnetic cascades in pulsars. *Astrophysical Journal, Part 1, Vol. 252, Jan. 1, 1982, p. 337-347.*, *252*, 337–347.

Di Marco, A., La Monaca, F., Poutanen, J., Russell, T., D., Anitra, A., Farinelli, R., Mastroserio, G., Muleri, F., Xie, F., Bachrtti, M., and Burderi, L. (2023). First detection of X-ray polarization from the accreting neutron star 4U 1820− 303. *The Astrophysical journal letters,* 953, no. 2 : L22.

Dyks, J., & Rudak, B. (2003). Two-pole caustic model for high-energy light curves of pulsars. *The Astrophysical Journal*, *598*(2), 1201.

Edlund. (1860). Über die Polarisation des Lichtes der Corona bei totalen Sonnenfinsternissen. Von Herrn Prof. Edlund in Stockholm. *Astronomische Nachrichten*, *52*(20), 305. https://doi.org/10.1002/asna.18600522001

Eichler, D., & Levinson, A. (2003). Polarization of Gamma-Ray Bursts via Scattering off a Relativistic Sheath. *The Astrophysical Journal*, *596*(2), L147. https://doi.org/10.1086/379313

Emslie, A. G., & Brown, J. C. (1980). The polarization and directivity of solar-flare hard X-ray bremsstrahlung from a thermal source. *Astrophysical Journal, Part 1, Vol. 237, May 1, 1980, p. 1015-1023.*, *237*, 1015–1023.

Fabiani, S., Capitanio, F., Iaria, R., Poutanen, J., Gnarini, A., Ursini, F., Farinelli, R., Bobrikova, A., Steiner, J. F., & Svoboda, J. (2024). Discovery of a variable energy-dependent X-ray polarization in the accreting neutron star GX 5- 1. *Astronomy & Astrophysics*, *684*, A137.

Farinelli, R., Fabiani, S., Poutanen, J., Ursini, F., Ferrigno, C., Bianchi, S., Cocchi, M., Capitanio, F., De Rosa, A., Gnarini, A., Kislat, F., Matt, G., Mikusincova, R., Muleri, F., Agudo, I., Antonelli, L. A., Bachetti, M., Baldini, L., Baumgartner, W. H., … Zane, S. (2023). Accretion geometry of the neutron star low mass X-ray binary Cyg X-2 from X-ray polarization measurements. *Monthly Notices of the Royal Astronomical Society*, *519*, 3681–3690. https://doi.org/10.1093/mnras/stac3726

Forsblom, S. V., Tsygankov, S. S., Poutanen, J., Doroshenko, V., Mushtukov, A. A., Ng, M., Ravi, S., Marshall, H. L., Di Marco, A., & La Monaca, F. (2024). Probing the polarized emission from SMC X-1: The brightest X-ray pulsar observed by IXPE. *Astronomy & Astrophysics*, *691*, A216.

Garg, A., Rawat, D., Bhargava, Y., Méndez, M., & Bhattacharyya, S. (2023). Flux-resolved spectropolarimetric evolution of the X-ray pulsar Hercules X-1 using IXPE. *The Astrophysical Journal Letters*, *948*(1), L10.

Garg, A., Rawat, D., & Méndez, M. (2024). Unveiling the X-ray polarimetric properties of LMC X- 3 with IXPE, NICER, and Swift/XRT. *Monthly Notices of the Royal Astronomical Society*, *531*(1), 585–591.

Goldstein, D. H. (2017). *Polarized Light* (3rd ed.). CRC Press. https://doi.org/10.1201/b10436

Gonzalez-Caniulef, D., Heyl, J., Fabiani, S., Soffitta, P., Costa, E., Bucciantini, N., Kirmizibayrak, D., & Xie, F. (2024). Crab pulsar: IXPE observations reveal unified polarization properties in the optical and soft X-ray bands. *Astronomy & Astrophysics*. https://doi.org/10.1051/0004-6361/202451815

Goosmann, R. W., & Matt, G. (2011). Spotting the misaligned outflows in NGC 1068 using X-ray polarimetry. *Monthly Notices of the Royal Astronomical Society*, *415*(4), 3119–3128.

Granot, J., & Königl, A. (2003). Linear polarization in gamma-ray bursts: The case for an ordered magnetic field. *The Astrophysical Journal*, *594*(2), L83.

Ingram, A., Ewing, M., Marinucci, A., Tagliacozzo, D., Rosario, D. J., Veledina, A., Kim, D. E., Marin, F., Bianchi, S., Poutanen, J., Matt, G., Marshall, H. L., Ursini, F., De Rosa, A., Petrucci, P.-O., Madejski, G., Barnouin, T., Gesu, L. D., Dovčiak, M., … Zane, S. (2023). The X-ray polarization of the Seyfert 1 galaxy IC 4329A. *Monthly Notices of the Royal Astronomical Society*, *525*(4), 5437–5449. https://doi.org/10.1093/mnras/stad2625

Jeffrey, N. L. S., & Kontar, E. P. (2011). Spatially resolved hard X-ray polarization in solar flares: Effects of Compton scattering and bremsstrahlung. *Astronomy & Astrophysics*, *536*, A93.

Kouch, P. M., Liodakis, I., Fenu, F., Zhang, H., Boula, S., Middei, R., Gesu, L. D., Paraschos, G. F., Agudo, I., Jorstad, S. G., Lindfors, E., Marscher, A. P., Krawczynski, H., Negro, M., Hu, K., Kim, D. E., Cavazzuti, E., Errando, M., Blinov, D., … Zane, S. (2024). *IXPE Observation of the Low-Synchrotron Peaked Blazar S4 0954+65 During An Optical-X-ray Flare* (arXiv:2411.16868). arXiv. https://doi.org/10.48550/arXiv.2411.16868

Kouch, P. M., Liodakis, I., Middei, R., Kim, D. E., Tavecchio, F., Marscher, A. P., Marshall, H. L., Ehlert, S. R., Gesu, L. D., Jorstad, S. G., Agudo, I., Madejski, G. M., Romani, R. W., Errando, M., Lindfors, E., Nilsson, K., Toppari, E., Potter, S. B., Imazawa, R., … Zane, S. (2024). IXPE observation of PKS 2155–304 reveals the most highly polarized blazar. *Astronomy & Astrophysics*, *689*, A119. https://doi.org/10.1051/0004-6361/202449166

Krawczynski, H., Garson III, A., Guo, Q., Baring, M. G., Ghosh, P., Beilicke, M., & Lee, K. (2011). Scientific prospects for hard X-ray polarimetry. *Astroparticle Physics*, *34*(7), 550–567.

Kushwaha, A., Jayasurya, K. M., Agrawal, V. K., & Nandi, A. (2023). IXPE and NICER view of black hole X-ray binary 4U 1630–47: First significant detection of polarized emission in thermal state. *Monthly Notices of the Royal Astronomical Society: Letters*, *524*(1), L15–L20.

Laha, S., Ricci, C., Mather, J. C., Behar, E., Gallo, L., Marin, F., Mbarek, R., & Hankla, A. (2025). X-ray properties of coronal emission in radio quiet active galactic nuclei. *Frontiers in Astronomy and Space Sciences*, *11*, 1530392. https://doi.org/10.3389/fspas.2024.1530392

Lei, F., Dean, A. J., & Hills, G. L. (1997). Compton Polarimetry in Gamma-Ray Astronomy. *Space Science Reviews*, *82*(3), 309–388. https://doi.org/10.1023/A:1005027107614

Majumder, S., Chatterjee, R., Jayasurya, K. M., Das, S., & Nandi, A. (2024). First Detection of X-Ray Polarization in Galactic Ultraluminous X-Ray Pulsar Swift J0243. 6+ 6124 with IXPE. *The Astrophysical Journal Letters*, *971*(1), L21.

Majumder, S., Kushwaha, A., Das, S., & Nandi, A. (2024). First detection of X-ray polarization in thermal state of LMC X-3: Spectro-polarimetric study with IXPE. *Monthly Notices of the Royal Astronomical Society: Letters*, *527*(1), L76–L81.

Marinucci, A., Muleri, F., Dovciak, M., Bianchi, S., Marin, F., Matt, G., Ursini, F., Middei, R., Marshall, H. L., Baldini, L., Barnouin, T., Rodriguez, N. C., De Rosa, A., Di Gesu, L., Harper, D., Ingram, A., Karas, V., Krawczynski, H., Madejski, G., … Zane, S. (2022). Polarization constraints on the X-ray corona in Seyfert Galaxies: MCG-05-23-16. *Monthly Notices of the Royal Astronomical Society*, *516*, 5907–5913. https://doi.org/10.1093/mnras/stac2634

Marra, L., Brigitte, M., Cavero, N. R., Chun, S., Steiner, J. F., Dovčiak, M., Nowak, M., Bianchi, S., Capitanio, F., & Ingram, A. (2024). IXPE observation confirms a high spin in the accreting black hole 4U 1957+ 115. *Astronomy & Astrophysics*, *684*, A95.

Middei, R., Perri, M., Puccetti, S., Liodakis, I., Di Gesu, L., Marscher, A. P., Cavero, N. R., Tavecchio, F., Donnarumma, I., & Laurenti, M. (2023). IXPE and Multiwavelength Observations of Blazar PG 1553+ 113 Reveal an Orphan Optical Polarization Swing. *The Astrophysical Journal Letters*, *953*(2), L28.

O'Dell, S. L., Attinà, P., Baldini, L., Barbanera, M., Baumgartner, W. H., Bellazzini, R., Bladt, J., Bongiorno, S. D., Brez, A., Cavazzuti, E., Citraro, S., Costa, E., Deininger, W. D., Del Monte, E., Dietz, K. L., Di Lalla, N., Donnarumma, I., Elsner, R. F., Fabiani, S., … Welch, D. (2019). The Imaging X-Ray Polarimetry Explorer (IXPE): Technical overview II. *UV, X-Ray, and Gamma-Ray Space Instrumentation for Astronomy XXI*, *11118*, 111180V. https://doi.org/10.1117/12.2530646

Paul, B. (2022). *The X-ray Polarimetry Satellite XPoSat*. *44*, 1853. 44th COSPAR Scientific Assembly. Held 16-24 July.

Ramsey, B. D., Bongiorno, S. D., Kolodziejczak, J. J., Kilaru, K., Alexander, C., Baumgartner, W. H., Elsner, R. F., McCracken, J., Mitsuishi, I., & Pavelitz, S. D. (2020). IXPE mirror module assemblies. *Optics for EUV, X-Ray, and Gamma-Ray Astronomy IX*, *11119*, 17–24. https://www.spiedigitallibrary.org/conference-proceedings-of-spie/11119/1111903/IXPE-mirror-module-assemblies/10.1117/12.2531956.short

Rawat, D., Garg, A., & Méndez, M. (2023a). Detection of X-Ray Polarized Emission and Accretion-disk Winds with IXPE and NICER in the Black Hole X-Ray Binary 4U 1630- 47. *The Astrophysical Journal Letters*, *949*(2), L43.

Rawat, D., Garg, A., & Méndez, M. (2023b). Spectropolarimetric study of 4U 1630- 47 in steep power-law state with IXPE and NICER. *Monthly Notices of the Royal Astronomical Society*, *525*(1), 661–666.

Rybicki, G. B., & Lightman, A. P. (1985). Basic Theory of Radiation Fields. In *Radiative Processes in Astrophysics* (pp. 51–76). John Wiley & Sons, Ltd. https://doi.org/10.1002/9783527618170.ch2

Saade, L., Kaaret, P., Bianchi, S., Gnarini, A., La Monaca, F., Ursini, F., Capitanio, F., Bobrikova, A., Di Marco, A., Heyl, J., Zane, S., Poutanen, J., & Collaboration, I. (2024). A First Look at the X-ray Polarization of 4U 1624-49 with IXPE. *American Astronomical Society, AAS Meeting #243, Id. 126.05. <ALTJOURNAL>Bulletin of the American Astronomical Society, Vol. 56, No. 2 e-Id 2024n2i126p05</ALTJOURNAL>*, *243*, 126.05.

Schnittman, J. D., & Krolik, J. H. (2010). X-ray polarization from accreting black holes: Coronal emission. *The Astrophysical Journal*, *712*(2), 908.

Secchi, P. A. (1860). Letter from Father Secchi to the Astronomer Royal on the Polarisation of Light reflected by the Moon. *Monthly Notices of the Royal Astronomical Society, Vol. 20, p.70*, *20*, 70. https://doi.org/10.1093/mnras/20.3.70

Soffitta, P. (1997). X-ray polarimetry an «almost» new frontier for X-ray astronomy. *Ital. Phys. Soc. Conf. Ser. 57: Frontier Objects in Astrophysics and Particle Physics*, 561. https://adsabs.harvard.edu/full/record/conf/foap./1997/1997foap.conf..561S.html

Soffitta, P. (2024). The Imaging X-ray Polarimetry Explorer (IXPE) and New Directions for the Future. *Instruments*, *8*(2), 25.

Soffitta, P., Attinà, P., Baldini, L., Barbanera, M., Baumgartner, W. H., Bellazzini, R., Bladt, J., Bongiorno, S. D., Brez, A., Castellano, S., Carpentiero, R., Castronuovo, M., Cavalli, L., Cavazzuti, E., D'Amico, F., Citraro, S., Costa, E., Deininger, W. D., D'Alba, E., … Seek, T. (2020). *The Imaging X-ray Polarimetry Explorer (IXPE): Technical overview III. 11444*, 1144462. Space Telescopes and Instrumentation 2020: Ultraviolet to Gamma Ray. https://doi.org/10.1117/12.2567001

Soffitta, P., Baldini, L., Bellazzini, R., Costa, E., Latronico, L., Muleri, F., Del Monte, E., Fabiani, S., Minuti, M., Pinchera, M., Sgro', C., Spandre, G., Trois, A., Amici, F., Andersson, H., Attina', P., Bachetti, M., Barbanera, M., Borotto, F., … Xie, F. (2021). The Instrument of the Imaging X-Ray Polarimetry Explorer. *The Astronomical Journal*, *162*(5), 208. https://doi.org/10.3847/1538-3881/ac19b0

Steiner, J. F., Nathan, E., Hu, K., Krawczynski, H., Dovciak, M., Veledina, A., Muleri, F., Svoboda, J., Alabarta, K., Parra, M., Bhargava, Y., Matt, G., Poutanen, J., Petrucci, P.-O., Tennant, A. F., Baglio, M. C., Baldini, L., Barnier, S., Bhattacharyya, S., … Xie, F. (2024). *An IXPE-Led X-ray Spectro-Polarimetric Campaign on the Soft State of Cygnus X-1: X-ray Polarimetric Evidence for Strong Gravitational Lensing* (arXiv:2406.12014). arXiv. https://doi.org/10.48550/arXiv.2406.12014

Stokes, G. G. (1851). On the composition and resolution of streams of polarized light from different sources. *Transactions of the Cambridge Philosophical Society*, *9*, 399.

Trippe, S. (2014). Polarization and Polarimetry: A Review. *Journal of Korean Astronomical Society*, *47*(1), 15–39. https://doi.org/10.5303/JKAS.2014.47.1.15

Ursini, F., Gnarini, A., Capitanio, F., Bobrikova, A., Cocchi, M., Di Marco, A., Fabiani, S., Farinelli, R., La Monaca, F., & Rankin, J. (2024). The IXPE View of Neutron Star Low-Mass X-ray Binaries. *Galaxies*, *12*(4), 43.

Veledina, A., Muleri, F., Poutanen, J., Podgornỳ, J., Dovčiak, M., Capitanio, F., Churazov, E., De Rosa, A., Di Marco, A., & Forsblom, S. V. (2024). Cygnus X-3 revealed as a Galactic ultraluminous X-ray source by IXPE. *Nature Astronomy*, 1–16.

Vivekanand, M. (2024). Phase-resolved Deadtime of the Crab Pulsar Using IXPE Data. *The Astrophysical Journal*, *972*(1), 36.

Waxman, E. (2003). New direction for gamma-rays. *Nature*, *423*(6938), 388–389. https://doi.org/10.1038/423388a

Weiskopf, M. C., Elsner, R. F., Hanna, D., Kapsi, V. M., Odell, S. L., Pavlov, G. G., & Ramsey, B. D. (2006). The prospects for X-ray polarimetry and its potential use for understanding neutron stars (2006). *arXiv Preprint Astro-Ph/0611483*.

Weisskopf, M. C., Silver, E. H., Kestenbaum, H. L., Long, K. S., & Novick, R. (1978). A precision measurement of the X-ray polarization of the Crab Nebula without pulsar contamination. *Astrophysical Journal, Part 2-Letters to the Editor, Vol. 220, Mar. 15, 1978, p. L117-L121.*, *220*, L117–L121.

West, E. A., & Smith, M. H. (1994). Polarization characteristics of the NASA Marshall Space Flight Ctr.(MSFC) Experimental Vector Magnetograph. *Polarization Analysis and Measurement II*, *2265*, 272–283. https://www.spiedigitallibrary.org/conference-proceedings-of-spie/2265/0000/Polarization-characteristics-of-the-NASA-Marshall-Space-Flight-Ctr-MSFC/10.1117/12.186699.short

Wilson, T. L., Rohlfs, K., & Hüttemeister, S. (2013). *Tools of Radio Astronomy*. Springer. https://doi.org/10.1007/978-3-642-39950-3